\documentstyle[aps,epsf]{revtex}
\newcommand{\be}{\begin{equation}}
\newcommand{\ee}{\end{equation}}
\def\reff#1{(\ref{#1})}

\newcommand{\1}{1\!\!\!\bot}
\newcommand{\ba}{\begin{eqnarray}}
\newcommand{\ea}{\end{eqnarray}}
\newcommand{\tr}{\mathop{\rm Tr}\nolimits}
\def\spose#1{\hbox to 0pt{#1\hss}}
\def\ltapprox{\mathrel{\spose{\lower 3pt\hbox{$\mathchar"218$}}
 \raise 2.0pt\hbox{$\mathchar"13C$}}}
\def\gtapprox{\mathrel{\spose{\lower 3pt\hbox{$\mathchar"218$}}
 \raise 2.0pt\hbox{$\mathchar"13E$}}}
\begin{document}
\draft
 
\renewcommand{\baselinestretch}{1.0} \large\normalsize

\title{A multigrid implementation of the Fourier acceleration method \\
       for Landau gauge fixing}
 
\author{Attilio Cucchieri$^{a,}$\thanks{E-mail address:
         cucchieri@roma2.infn.it}
and Tereza Mendes$^{a,b,}$\thanks{E-mail address:
              mendes@roma2.infn.it}}
\address{{\small $^a$Gruppo APE, Dipartimento di Fisica,
                  Universit\`a di Roma ``Tor Vergata''}         \\
  {\small $^b$INFN, Sezione di Roma 2}  \\
  {\small Via della Ricerca Scientifica 1, 00133 Roma, ITALY}}

\date{February 12, 1998}
\maketitle
\begin{abstract}
We present a new implementation of the Fourier acceleration 
method for Landau gauge fixing.
By means of a multigrid inversion we are able to avoid the
use of the fast Fourier transform. This makes the method
more flexible, and well suited for vector and parallel machines.
We study the performance of this algorithm on serial and on
parallel (APE100) machines for the 4-dimensional $SU(2)$ case. 
We find that our method is equivalent to the standard implementation
of Fourier acceleration already on a serial machine, and that
it parallelizes very efficiently: 
its computational cost shows a linear speedup with the number of 
processors.
We have also implemented, on the parallel machines, a version of the
method using conjugate gradient instead of multigrid. This leads to an 
algorithm that is efficient at intermediate lattice volumes.
\end{abstract}


\section{Introduction}

Lattice gauge fixing is a necessary tool for understanding 
the relationship between continuum and lattice gauge theory.
In fact, due to asymptotic freedom, the continuum limit of
the lattice theory is the weak-coupling limit, and a weak-coupling
expansion requires gauge fixing.
Gauge fixing is also used in smearing techniques,
and is necessary in order to evaluate quark/gluon matrix 
elements which can be used to extract
non-perturbative results from Monte Carlo simulations \cite{Rossi}.
It is therefore important to devise numerical algorithms
to efficiently gauge fix a lattice configuration.
An important issue regarding the efficiency of these 
algorithms is the problem of {\em critical slowing-down} (CSD),
which occurs when the relaxation time $\tau$ of an algorithm 
diverges as the lattice volume is increased \cite{CSD}. Conventional
local algorithms have {\em dynamic critical exponent} $z\approx 2$,
namely $\tau$ grows with the lattice side $N$ roughly as $N^2$.
Improved local methods show typically $z \approx 1$, while
global methods may succeed in eliminating CSD completely,
i.e.\ $z\approx 0$. Usually, global algorithms are
more costly per iteration than
local methods but, due to the elimination of CSD,
their total computational cost becomes
progressively lower than that of local methods
at large lattice volumes. For this reason, efficient
global algorithms are a highly desirable tool
in large-volume applications. Another important issue is
whether gauge-fixing algorithms can be implemented efficiently
on parallel machines \cite{SS}, since computers of this type
are widely used in numerical studies of gauge theory on large 
lattices. 

A well-known global approach for reducing CSD, applicable to
gauge fixing as well as to other problems, is the method of
{\em Fourier acceleration} (FA)
\cite{Fourier}. The idea is to precondition a problem
using a diagonal matrix in momentum space which is related
to the solution of a simplified version of the problem
\cite{Fourier,AllF}. For the $SU(2)$ Landau gauge fixing case it 
can be proved \cite{gfix23} that
Fourier acceleration eliminates CSD completely at infinite $\beta$,
namely the dynamic critical exponent $z$ is equal to
zero; we have also obtained \cite{gfix1}, at finite $\beta$
and in two dimensions, that $z$ is equal to $0.036 \pm 0.064$.
Moreover, the FA gauge-fixing method is very efficient in
achieving a constant value for the longitudinal gluon propagator at
zero three-momentum \cite{gfix23,gfix1,Mare}, which provides a very
sensitive test of the goodness of the gauge fixing.

To fix Landau gauge on the lattice one looks for a minimum of the
functional \cite{W2}
\be
{\cal E}_{U}[ g ] \equiv
\frac{1}{4\,V} \sum_{\mu\mbox{,}\, x} \, \tr \,
   \left[ \, \1 \; - \; g(x)\,U_{\mu}(x)\,g^{\dagger}(x+e_{\mu})\,
   \right]
\label{eq:Etomin}
\;\mbox{.}
\ee
(We refer to \cite{gfix1} for notation.) The FA update is given by
$g^{(new)}(x) \equiv R(x) \, g^{(old)}(x)$ with
\be
R(x) \, \propto \, \1 - {\widehat F}^{-1}\left[
         \, \frac{\alpha}{p^{2}(k)} \,
                {\widehat F}
 \left( \nabla \cdot A^{\left( g \right)} \right)
\right](x)
\;\mbox{.}
\label{update}
\ee
Here ${\widehat F}$ is a Fourier transform,
$\alpha$ is a tuning parameter,
$p^{2}(k)$ is the square of the lattice momentum, and
$\nabla \cdot A$ is the lattice divergence of the
gluon field $A_{\mu}$. Thus, in this case, the preconditioning is
obtained using in momentum space a diagonal matrix
with elements given by $1 / p^{2}(k)$ (see
\cite[Section II]{Fourier} for details).
The FA method in its original form makes use of the 
{\em fast Fourier transform} (FFT) to evaluate ${\widehat F}$
and ${\widehat F}^{-1}$, which requires a work
of only $V\log V$ where $V$ is the lattice volume \cite{Fourier}, making it 
very appealing from the computational point of view.
On the other hand, in order to implement FFT efficiently, one
is restricted to using lattice sides that are powers of 2 
\cite[Chapter 12]{PTVF}. Moreover,
implementing FFT on parallel machines of the SIMD
type, and especially in 4 dimensions, can be very cumbersome
\cite{SS,foot1}. 
Here we present a new implementation of the FA method for Landau 
gauge fixing which avoids completely
the use of the fast Fourier transform, and we test its
performance for the $4$-dimensional $SU(2)$ case on serial
and on parallel machines. Preliminary results
have been reported in \cite{lat97}.

\section{The multigrid FA method}

Let us start by noting
that the Fourier-mode decomposition in eq.\ \reff{update}
is equivalent to an inversion of the lattice Laplacian operator $\Delta$:
\be
{\widehat F}^{-1}\,p^{-2}(k)\,
{\widehat F}
\;=\;\left(\,- \Delta\,\right)^{-1} \;\mbox{.}
\ee
(Note that this inversion is carried out for each component of 
$\nabla\cdot A\,$.)
Thus, the FFT can be avoided by inverting $\Delta$ using an alternative
algorithm. Clearly, a good candidate should be suitable for use on
parallel machines, and should require, ideally, the same computational work
as FFT, i.e.\ $V\log V$. One such candidate is the {\em multigrid} (MG)
algorithm with W-cycle and piecewise-constant interpolation.
The idea of MG is to solve the lattice problem recursively,
using local (Gauss-Seidel) updates on coarser versions of 
the original lattice in order to accelerate
the convergence of slow modes of the solution.
MG is an efficient {\em iterative} routine for inverting the Laplacian $\Delta$:
with our choice of cycle and interpolation, each iteration
of the method represents a work of order $V$, and the number of iterations
required for convergence is proportional to $\log V$ at most \cite{MG,foot0}. 
Moreover, the MG routine can be successfully
implemented on vector and parallel machines \cite{parallel_MG}.
Thus this approach should preserve the property of eliminating 
CSD for Landau gauge fixing, while being applicable 
on parallel machines.
At the same time, there is no restriction on the lattice
size since, even for a fixed coarsening factor (e.g.\ 2),
the size of the coarsest grid can be adjusted conveniently.

The overhead for the MG routine is likely to be larger than the
one for FFT, but in our case it can be reduced by exploiting 
the fact that multigrid (as opposed to FFT) is an iterative method.
For example, a significant computational gain can be obtained
if one starts the inversion from a good initial guess for 
the solution. Also, by changing the stopping criterion for the
inversion, the accuracy of the solution
can be suitably varied, while for FFT the accuracy is fixed
by the precision used in the numerical code. This is important
since we will be tuning the parameter $\alpha$ in eq.\ \reff{update},
and this tuning can be done only up to an accuracy of a few percent.
Thus, the inversion of $\Delta$ most likely will not require
the high accuracy obtained in the FFT case, making possible a
substantial reduction of the computational cost.

In order to test the feasibility of this approach
we have started our simulations
on a workstation, comparing the performance
of the MG implementation of the Fourier acceleration method (MGFA)
with the original implementation using FFT (which we call from now 
on FFTFA) \cite{single-prec}.
As a first step, we tuned the parameter $\alpha$ for the
FFTFA method at infinite $\beta$ and $V= 8^4$. We obtained
$\alpha_{opt} = 1.105$ as optimal choice, and a number of
gauge-fixing sweeps equal to $14.9(2)$.
(Note that our data represent averages over a set of gauge
configurations, and that our error bars are one standard
deviation.)
We then tested MGFA with $\alpha = \alpha_{opt}$:
in addition to the W-cycle (where each
grid is visited twice before proceeding to the next finer grid),
we also tested for comparison the V-cycle (where each grid is
visited once before proceeding to the next finer grid) and the
standard Gauss-Seidel update; for the W-cycle
we also varied the number of Gauss-Seidel relaxation sweeps on 
each grid ($N_{r}$), and the minimum number of complete multigrid
cycles ($N_{min}$). Results for the number of gauge-fixing sweeps
as a function of the accuracy \cite{foot2}, at infinite $\beta$ 
and $V= 8^4$, are reported in Fig.\ \ref{FIG:accuracy}.
By comparing these results with the result from the FFTFA method,
it is clear that the number of gauge-fixing sweeps is
independent of the method used for the inversion of $\Delta$,
provided that a high enough accuracy is required. Among
the tested versions of the MGFA method, the best from the
point of view of computational cost is the one with the following
choice of parameters: $\gamma = 2$, an accuracy of about $10^{-5}$,
two relaxation sweeps on each grid, and a minimum of two multigrid cycles
for each inversion of $\Delta$. (We note that the CPU cost per
iteration of this MG version is higher than for the other versions,
but the fact that the inversion of the Laplacian can be stopped
already at an accuracy of $10^{-5}$ makes it the fastest version.)
{\em We then adopt this version of MG as the routine used for our
MGFA method.} For the MG routine we have chosen to use 
a coarsest grid of $2^4$. Nevertheless we have checked
that the performance of the chosen version does not change if a coarsest
grid of $4^4$ is used. This is an important result, as we will see, for the
implementation of the algorithm on parallel machines.
We have also tested different initial guesses for the MG inversion,
finding that the use of the solution to
the previous inversion of the Laplacian is preferable to
a null or random initial guess.

\section{Results and Conclusions}

Initially, we have tested the performance of the MGFA method at $\beta
= \infty$ and at $\beta = 2.2$ on a workstation. The results, reported in
Table \ref{tab:accuracy}, are compared
with those obtained for the FFTFA method and for
the overrelaxation (OVE) algorithm \cite{MO3},
an improved local method which shows $z\approx 1$
\cite{gfix23,gfix1} and which is often used for production runs
in lattice gauge theory.
In all cases the stopping criterion for the gauge fixing
was $(\nabla\cdot A)^2 \leq 10^{-12}$.
From the data at $\beta = \infty$ we can study the volume
dependence of the computational cost for the various
algorithms.
Clearly, FFTFA and MGFA have a similar performance \cite{foot3},
showing a number of gauge-fixing sweeps increasing less than
logarithmically with the volume $V$,
and the CPU time per sweep increasing roughly as $V \log V$.
From our data we note that the number of MG cycles per inversion is
essentially independent of the volume.
As expected, for the overrelaxation algorithm the
number of gauge-fixing sweeps is proportional to the lattice side $N$,
and the CPU time per sweep grows as the volume $N^4$.
The data for the total CPU time for the two implementations of
the FA method and for the overrelaxation method are well fit
by $V\log V$ and $N^5$, respectively.
An extrapolation of our data
using these fits predicts that either
version of the FA method would be less costly than
the overrelaxation method already at lattice
side $N=24$.
Actually, the CPU cost for MGFA scales slightly better
with the volume than for FFTFA. Thus, on a serial machine,
and at $\beta = \infty$,  the fast Fourier transform can be
successfully replaced by the MG routine, and MGFA should
be the method of choice in the limit of
large lattice volume.
MGFA is equivalent to FFTFA also at $\beta = 2.2$ and $V = 8^4$,
and therefore it appears that the use of FFT can be avoided also 
at finite $\beta$.

We remark that the FA method may have convergence problems at low
values of $\beta$ \cite{gfix23,Gutbrod}, probably related to 
the large number of local minima of the functional ${\cal E}_{U}[ g ]$
and/or to the existence of topological objects on the lattice.
These problems are more likely to affect a global method than a local
one (such as overrelaxation). A possible solution may be 
the smearing approach recently introduced in \cite{Hetrick}: by 
smoothing out the lattice gauge configuration, one can perform gauge fixing 
at an effective $\beta$ close to infinity; this
result is then used as a preconditioning of the original gauge-fixing 
problem, i.e.\ the gauge-fixing iterations start already close to a minimum. 
This approach aims at eliminating the problem of Gribov copies on the
lattice, and is very well suited for the FA method.
In fact, the two gauge-fixing steps involved (the one for the 
preconditioning --- at high $\beta$ --- and the one starting close to 
a minimum) are ideal applications of FA.

\vskip 3mm

We then applied MGFA on two parallel (APE-Quadrics) machines of the
APE100 series: the Q1 ($2^3$ processors) and the QH4 
($8^3$ processors) \cite{foot4}.
In order to implement the method on a parallel machine,
the idea is to use as the coarsest grid for the
MG routine a grid with volume equal to or
larger than the number of processors.
This avoids the problem of idle processors discussed
in ref.\ \cite{parallel_MG}.
For example, for the Q1 we implemented the MG routine with coarsest
grid $4^4$ (respectively $6^4$) for the lattice volume $V = 8^4$
(respectively $V = 12^4$), while for the QH4 we used $8^4$ as the 
coarsest grid \cite{MG+CG}.
We have checked that the number of
gauge-fixing sweeps does not change if we use for the MG routine
an accuracy in the range $10^{-4}$--$10^{-5}$. This
confirms the result obtained for a serial 
machine (see Fig.\ \ref{FIG:accuracy}).
In all our runs on APE machines we set the accuracy for the
inversion to $5 \times 10^{-5}$.
Since these machines work in single precision
we have also decreased the stopping criterion for the
gauge fixing to $(\nabla\cdot A)^2 \leq 10^{-7}$.

As an alternative to the MG inversion,
we have also implemented a version of the Fourier acceleration method in
which the Laplacian $\Delta$ is inverted using a standard (iterative)
conjugate-gradient (CG) method. We call this method CGFA.
The CG algorithm is simpler to program, and has been widely used on
parallel machines.
Its CPU time per cycle should be smaller than the one for MG, but
the number of iterations required in order
to achieve a fixed accuracy should increase faster than for
the MG routine. 
In fact, we have checked that the number of multigrid iterations
is essentially independent of the lattice volume,
while the number of CG iterations grows roughly as $N^{0.37}$.

In Table \ref{tab:cubetto} we report
the number of gauge-fixing sweeps obtained,
at $\beta = \infty$ and for several lattice sizes,
for the MGFA and the CGFA methods on APE machines. Their
performances are compared with that of a standard overrelaxation (OVE)
and with that of an unaccelerated local algorithm (the so-called
Los Alamos algorithm, LOS) \cite{gfix23,gfix1}.
These runs on parallel machines confirm that local algorithms
are usually not able to achieve a constant value for the
longitudinal gluon propagator at zero three-momentum. This can be
checked by changing the stopping criterion for the gauge fixing:
instead of considering $(\nabla\cdot A)^2$ we can consider
the quantity $e_6$ defined in \cite{gfix23,gfix1}, which 
monitors the fluctuations of this gluon propagator. Results are
reported in Table \ref{tab:cubetto} for the lattice volume
$V = 32^4$. (Note that for LOS and OVE
the results, in this case, have a large statistical error, due to the
fact that for some configurations the gauge fixing did not converge
within $3000$ sweeps.)

Also in the parallel case, the FA method eliminates CSD at $\beta=\infty$,
and therefore should be the method of choice on large lattices.
With respect to the overrelaxation method, the two implementations of FA
are competitive already at volume $32^4$ if one requires a sensitive  
stopping criterion for the gauge fixing. We observe that, at our lattice 
sizes, the CG implementation is about two times faster than MGFA, but
at large volumes we expect MGFA to win out.

We recall that local 
methods are more efficiently implemented on parallel machines than
global ones, since they require smaller communication between processors.
Global methods need implementations
specifically designed for parallel machines in order to
achieve a significant reduction of their computational overhead.
For example, in a parallel implementation, the update for MG is 
not exactly of the Gauss-Seidel type, 
and in fact we observe that a higher (fixed) number of 
MG iterations is needed \cite{Jacobi}.
We think that our code for the MG routine can 
be optimized, since we have not explored more advanced features of MG
that can play a role on parallel machines, such as asynchronous 
multigrid and the use of accommodative cycles instead of a fixed
cycling strategy.

We have checked the dependence of the performance of the algorithms
on the number of processors using the data from the Q1 and from the QH4,
respectively for lattice volume $12^4$ and $32^4$. 
The CPU time per gauge-fixing iteration per site scales down by a factor
of approximately 62 for all the four algorithms, showing that the
two FA methods parallelize as efficiently as the local ones.
(Note that the number of processors increases by a factor 64 going 
from the Q1 to the QH4.)

\vskip 3mm

We have shown that the fast Fourier transform in the Landau 
gauge-fixing Fourier acceleration method can be successfully substituted ---
on serial as well as on parallel machines --- 
by an alternative inversion routine.
This idea can in
principle be extended to other applications of FA such as the case of
quark propagators, and the Monte Carlo method for thermalization of
lattice configurations.

\section*{Acknowledgements}

Simulations were done on
an IBM RS-6000/340 workstation at New York University, and
on two APE-Quadrics machines at the University of Rome Tor Vergata.
We wish to thank Philippe de Forcrand and Alan Sokal for helpful
discussions.


%
%

\begin{figure}[t]
\begin{center}
\vspace*{0cm} \hspace*{-0cm}
\epsfxsize=0.4\textwidth
\leavevmode\epsffile{gfsweeps_vs_acc.ps}
\hspace{2cm}
\epsfxsize=0.4\textwidth
\epsffile{gfsweeps_vs_acc_ZOOM.ps}  \\
\caption{~Plot of the number of gauge-fixing sweeps for the
MGFA algorithm as a function of the accuracy imposed on the inversion of the
Laplacian. Simulations were carried out on a workstation, for lattice volume
$8^4$ and at $\beta = \infty$. The different types
of multigrid cycle employed are: {\bf (a)} Gauss-Seidel update ($\Diamond$),
$V$-cycle with $N_r=1, N_{min}=1$ ($\bigcirc$), 
$W$-cycle with $N_r=1, N_{min}=1$ ($\Box$),
$W$-cycle with $N_r=2, N_{min}=1$ ($\ast$);
{\bf (b)} a closer view of the $W$-cycle with $N_r=2, N_{min}=1$ case
(solid line), together with the $W$-cycle with $N_r=2, N_{min}=2$
case ($\times$).
In the second plot errors are omitted for clarity.
We note that for FFTFA the number of gauge-fixing sweeps is
$14.9\pm 0.2$. In all cases we set $\alpha = 1.105$, and we stopped the gauge fixing
when the condition $(\nabla\cdot A)^2 \leq 10^{-10}$ was satisfied.}
\label{FIG:accuracy}
\end{center}
\end{figure}

\clearpage

%
%

\widetext
\begin{table}
\caption{Comparison between the FFTFA and MGFA algorithms at
$\beta = \infty$ and at $\beta = 2.2$ on a workstation. 
The optimal choice for the tuning parameter $\alpha$ for the
two FA methods is reported \protect\cite{tuning}.
We also quote here results for the overrelaxation (OVE) method.}
\narrowtext
\label{tab:accuracy}
\begin{tabular}{cccccc}
algorithm & $\beta$ & $ V $ & $\alpha_{opt}$ & GF sweeps & CPU time (s) \\ 
\hline
OVE   & $\infty$ & $ 4^4 $ &        & $18.1 \pm 0.1$ & $0.167 \pm 0.003$ \\ 
FFTFA & $\infty$ & $ 4^4 $ & $1.12$ & $13.6 \pm 0.3$ & $0.34 \pm 0.01$ \\ 
MGFA  & $\infty$ & $ 4^4 $ & $1.12$ & $13.7 \pm 0.3$ & $0.55 \pm 0.02$ \\
\hline
OVE   & $\infty$ & $ 8^4 $ &        & $37.5 \pm 0.2$ & $5.91 \pm 0.03$ \\
FFTFA & $\infty$ & $ 8^4 $ & $1.12$ & $16.1 \pm 0.3$ & $9.60 \pm 0.20$ \\
MGFA  & $\infty$ & $ 8^4 $ & $1.12$ & $16.2 \pm 0.3$ & $11.4 \pm 0.5$ \\
\hline
OVE   & $\infty$ &$ 16^4 $ &        &  $76.6 \pm 0.9$ & $194 \pm 2$ \\
FFTFA & $\infty$ &$ 16^4 $ & $1.1$ & $20.0 \pm 0.4$ & $246 \pm 6$ \\
MGFA  & $\infty$ &$ 16^4 $ & $1.1$ & $20.3 \pm 0.4$ & $242 \pm 10$ \\
\hline
OVE   & $2.2$ & $ 8^4 $ &        & $139 \pm 8$ & $22.5 \pm 1.4$ \\ 
FFTFA & $2.2$ & $ 8^4 $ & $1.89$ & $333 \pm 27$ & $200 \pm 12$ \\
MGFA  & $2.2$ & $ 8^4 $ & $1.89$ & $312 \pm 25$ & $187 \pm 15$
\end{tabular}
\end{table}
 
\widetext
\begin{table}
\caption{\label{tab:cubetto}
Results for the so-called Los Alamos (LOS) method,
the overrelaxation (OVE) method, the
CGFA and MGFA algorithms at
$\beta = \infty$.
Runs were carried out on the Q1 and QH4
machines, for the lattice volumes
$V = 8^4, 12^4$ and $V = 16^4, 32^4$ respectively.
The optimal choice for the tuning parameter $\alpha$ for the
two FA methods is also reported.
In all cases we stopped the gauge fixing
when the condition $(\nabla\cdot A)^2 \leq 10^{-7}$ was satisfied.
For the second set of data at $V = 32^4$ the condition
$e_{6} \leq 5 \times 10^{-6}$ has been used.}
\narrowtext
\begin{tabular}{ccccc}
algorithm & $ V $ & $\alpha_{opt}$ & GF sweeps & CPU time (s) \\ \hline
LOS  & $8^4$    &      &  57.3 $\pm$ 0.6 & $\approx 1$ \\ 
OVE  & $8^4$    &      &  23.0 $\pm$ 0.3 & $ < 1$ \\ 
CGFA & $8^4$    &$1.14$&  13.9 $\pm$ 0.1 & 4.56 $\pm$ 0.07 \\ 
MGFA & $8^4$    &$1.10$&  13.8 $\pm$ 0.1 & 13.4 $\pm$ 0.2 \\ 
\hline
LOS  & $12^4$   &      &  117 $\pm$ 2 & 18.4 $\pm$ 0.4 \\ 
OVE  & $12^4$   &      &  33.6 $\pm$ 0.5 & 5.9 $\pm$ 0.1 \\ 
CGFA & $12^4$   &$1.22$&  16.4 $\pm$ 0.3 & 34.1 $\pm$ 0.7 \\ 
MGFA & $12^4$   &$1.24$&  16.2 $\pm$ 1.0 & 92 $\pm$ 2 \\ 
\hline
LOS  & $16^4$ &        &  198 $\pm$ 2 & $\approx 1$ \\ 
OVE  & $16^4$ &        &  46.3 $\pm$ 0.6 & $ < 1$ \\
CGFA & $16^4$ & $1.33$ &  17.6 $\pm$ 0.2 & 2.12 $\pm$ 0.04 \\
MGFA & $16^4$ & $1.26$ &  17.3 $\pm$ 0.1 & 5.95 $\pm$ 0.08 \\
\hline
LOS  & $32^4$ &        &  640 $\pm$ 20 & 83 $\pm$ 2 \\ 
OVE  & $32^4$ &        &  84 $\pm$ 2 & 12.0 $\pm$ 0.3 \\ 
CGFA & $32^4$ & $1.38$ &  21.5 $\pm$ 0.7 & 49 $\pm$ 1 \\ 
MGFA & $32^4$ & $1.35$ &  20.7 $\pm$ 2.0 & 98 $\pm$ 3 \\ 
\hline 
LOS  & $32^4$ &        &  1970 $\pm$ 90 & 553 $\pm$ 30 \\
OVE  & $32^4$ &        &  252 $\pm$ 70 & 75 $\pm$ 20 \\
CGFA & $32^4$ & $1.34$ &  23.4 $\pm$ 0.8 & 57 $\pm$ 2 \\
MGFA & $32^4$ & $1.33$ &  22 $\pm$ 3 & 123 $\pm$ 3 
\end{tabular}
\end{table}


\begin{references}
\bibitem{single-prec} We have used single precision for the FFT and for the
             MG inversion routines, while for the gauge fixing we have
             used double precision.
\bibitem{foot2} Our definition of accuracy is $| r(t) | / | r(0) |$,
                where $r(t)$ is the residual at the $t$-th iteration.
\bibitem{tuning} The tuning of $\alpha$ was done independently
for the MGFA and the FFTFA method. In all cases the optimal choice
$\alpha_{opt}$ was the same. This result confirms
that the use of a different routine
for inverting the Laplacian $\Delta$ does not modify the performance
of the gauge-fixing algorithm, if a high enough accuracy is
required for this inversion.
\bibitem{Jacobi} We have checked that increasing the number of relaxation
                 sweeps per grid causes a reduction of the total MG cycles
                 needed per inversion, but does not lead to a computational 
                 gain.
\end{references}
\end{document}